\documentclass[fleqn,12pt,twoside]{article}
\usepackage{espcrc1}

\usepackage{graphicx}
\usepackage[figuresright]{rotating}

\newcommand{\AmS}{{\protect\the\textfont2
  A\kern-.1667em\lower.5ex\hbox{M}\kern-.125emS}}

\title{Evidence for the Jacobi shape transition in hot $^{46}$Ti}

\author{A.~Maj\address[IFJ]{The Niewodnicza\'nski Institute of Nuclear Physics,  
        Polish Academy of Siences, Radzikowskiego 152,
	PL-31342 Krak\'ow, Poland}, M.~Kmiecik\addressmark[IFJ], 
	A.~Bracco\address[MI]{Dipartimento di Fisica and INFN sez. Milano, 
	I-20133 Milano, Italy}, F.~Camera\addressmark[MI], 
	P.~Bednarczyk\addressmark[IFJ]\address[IRES]
	{Institut de Recherches Subatomiques, 23 rue du Loess, 
	F-67037 Strasbourg, France},
	B.~Herskind\address{The Niels Bohr Insitute, Blegdamsvej 17, 
	DK-2100 Copenhagen, Denmark},\\
        S.~Brambilla\addressmark[MI],
	G.~Benzoni\addressmark[MI], M.~Brekiesz\addressmark[IFJ],
        D.~Curien\addressmark[IRES], G.~De~Angelis\address[LNL]
        {INFN - Laboratori Nazionali di Legnaro, I-35020 Legnaro (PD), Italy},
        E.~Farnea\address[Pad]{INFN sez. Padova, I-35131 Padova, Italy}, 
	J.~Gr\c{e}bosz\addressmark[IFJ],
	M.~Kici\'nska-Habior\address{Institute 
	of Experimental Physics, Warsaw University, PL-00681 Warsaw, Poland},
        S.~Leoni\addressmark[MI], W.~M\c{e}czy\'nski\addressmark[IFJ], 
	B.~Million\addressmark[MI], D.R.~Napoli\addressmark[LNL],
	J.~Nyberg\address{Departament of Radiation Sciences,
	Uppsala University, SE-75121 Uppsala, Sweden}, 
	C.M.~Petrache\address{Dipartimento di Fisica, 
        Universit\'a di Camerino, I-62032 Camerino (MC) , Italy},
	J.~Stycze\'n\addressmark[IFJ],
        O.~Wieland\addressmark[MI], 
	M.~Zi\c{e}bli\'nski\addressmark[IFJ], K.~Zuber\addressmark[IFJ],
        N.~Dubray\addressmark[IRES], J.~Dudek\addressmark[IRES] 
	and K.~Pomorski\address[UMCS] {Uniwersytet Marii Curie-Sk{\l}odowskiej,
	PL-20031 Lublin, Poland}}

\begin{document}

\maketitle

\begin{abstract}
The $\gamma$-rays from the decay of the GDR in the compound nucleus reaction
$^{18}$O+$^{28}$Si at bombarding enery of 105~MeV have been
measured in an experiment using a setup consisting of the combined
EUROBALL IV, HECTOR and EUCLIDES arrays.
The shape of the rotating compound nucleus, $^{46}$Ti, is expected to undergo 
the Jacobi transition around spin 28~$\hbar$. A comparison of the GDR
lineshape data with the predictions of the thermal shape fluctuation model, 
based on the most recent rotating liquid drop
LSD calculations, shows evidence for such Jacobi shape transition.
In addition to the previously found broad structure in the GDR lineshape 
region at 18--25~MeV caused
by large deformations, the presence of a low energy component 
(around 10~MeV), due to the Coriolis splitting in prolate, well deformed 
shape has been identified for the first time.
\end{abstract}

\section{INTRODUCTION}

The Jacobi shape transition, an abrupt change of nuclear shape from an oblate 
ellipsoid  non-collectively rotating around its symmetry axis, to an elongated 
prolate or triaxial shape, rotating collectively around the shortest axis 
has been predicted to appear in many nuclei at angular momenta close 
to the fission limit~\cite{Swiat1,Dudek}.
Signatures of the presence of elongated shapes can be found,  
among others, in the $\gamma$-decay of the 
Giant Dipole Resonance (GDR), in the giant back-bend 
of the $E$2 $\gamma$-transition energies~\cite{Ward} and in the angular 
distribution of the emitted charged particles.

Until now, however, a firm experimental evidence for the Jacobi shape transition in nuclei 
could not be established. Some indications of such elongated shapes 
were obtained in the studies of the GDR decay 
from $^{45}$Sc$^*$~\cite{MKH} and $^{46}$Ti$^*$~\cite{Maj_Osaka} compound nuclei.
In the case of $^{45}$Sc$^*$, 
the GDR measurement was inclusive, while for $^{46}$Ti$^*$ it was more
exclusive, namely it was associated 
with various $\gamma$-multiplicity values. However, in both studies other 
explanations, as for example contribution from fission could not be 
completely ruled out.     

In order to address in more detail the problem of the Jacobi shape transition 
focusing on the GDR $\gamma$-decay, a new highly selective experiment
for $^{46}$Ti$^*$ was performed. To obtain high energy $\gamma$-ray spectra 
associated with fusion-evaporation reaction and free from contribution 
of fission products and/or other non-fusion reactions, the restrictive 
conditions of gating on well-known discrete 
lines of the evaporation residual nuclei were applied.

In this paper we report on some new results concerning the measurement 
of the high energy $\gamma$-rays with very stringent conditions 
selecting the highest part of the spin distribution.

The following section describes the experiment and the data analysis.
The discussion of the results presented here focuses for the first time 
on the importance of the Coriolis splitting of the GDR components at the high
rotational frequencies that can be reached only in light nuclei.

\section{THE EXPERIMENT}

The experiment was performed at the VIVITRON accelerator of the IReS Laboratory,
Strasbourg (France), using the EUROBALL~IV Ge-array coupled to the HECTOR 
array~\cite{Maj94} and the charged particle detector EUCLIDES. The $^{46}$Ti
compound nucleus was populated in the $^{18}$O+$^{28}$Si reaction  at 105~MeV
bombarding energy. The excitation energy of the $^{46}$Ti nuclei 
was  86~MeV and the maximum  angular momentum $l_{max}\approx$~35~$\hbar$. 
For this experiment, the EUROBALL consisted of 26 germanium clover 
and 15 cluster detectors (all with the BGO anti-Compton shields), 
and 75\% of the Inner-ball (83 BGO crystals) which together 
with the  germanium detectors resulted in 65\% efficiency 
for the multiplicity determination. The 8 large volume
BaF$_2$ detectors of the HECTOR were placed 
in the forward  hemisphere, together
with 4 small BaF$_2$ detectors which provided a good time reference signal. 
The EUCLIDES  consisted of 40 silicon telescope detectors  covering 
approximately 90\% of the solid angle. Only events having at least 
2 Compton suppressed Ge signals and one high-energy $\gamma$-ray were
accepted. 
A total number of 10$^8$ events were collected, in which the $\gamma$-ray 
energy in any BaF$_2$ detector was larger then 4~MeV. 
\begin{figure}[htb]
\begin{minipage}[t]{85mm}
\includegraphics[scale=0.9]{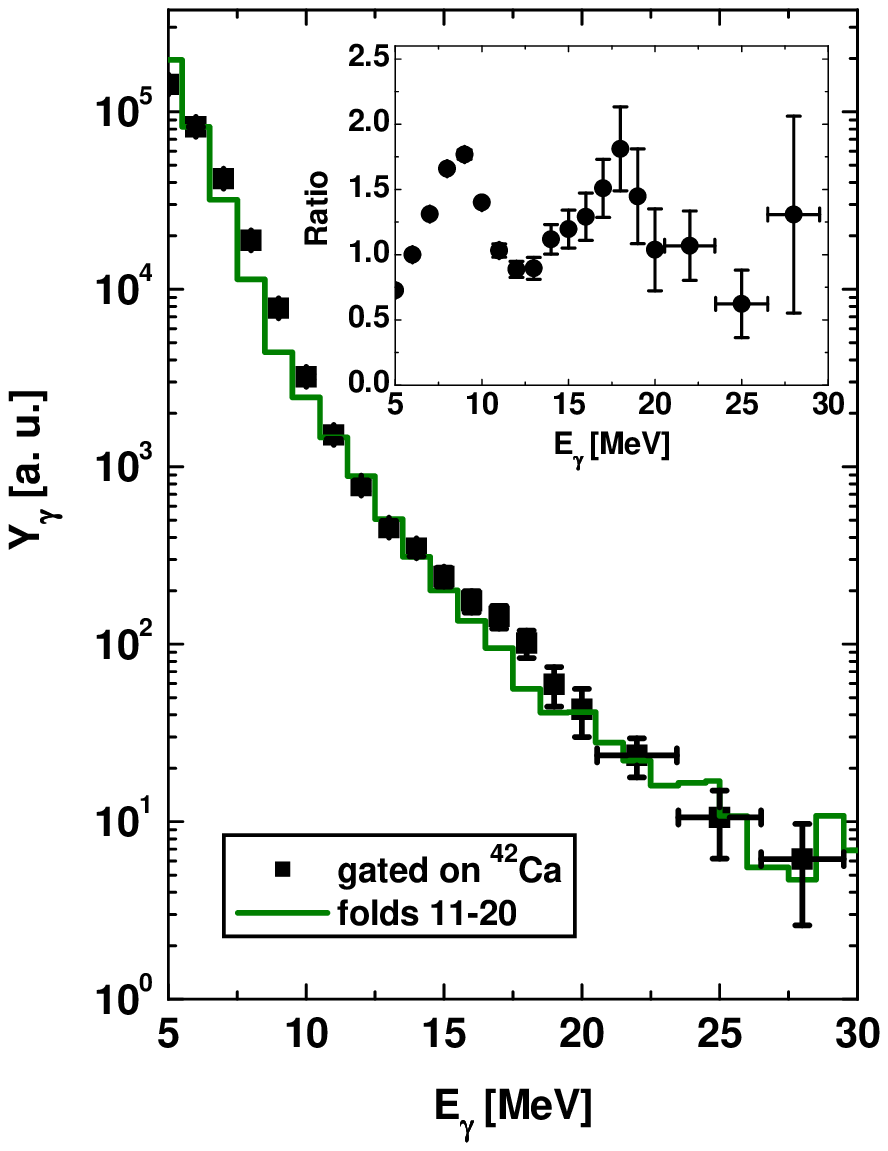}
\end{minipage}
\begin{minipage}[b]{72mm}
\caption{
Two high-energy $\gamma$-ray spectra measured with the HECTOR BaF$_2$-detectors:  
one is gated by the Innerball folds 11--20  (histogram) and
the other is additionally gated by the discrete transitions in $^{42}$Ca 
measured in the EUROBALL Ge-detectors (filled squares).
The inset shows the ratio of these two spectra.}
\label{fig:1}
\end{minipage}
\end{figure}

Figure~\ref{fig:1} shows two measured high-energy $\gamma$-ray spectra. 
The first, displayed by the histogram, corresponds to the selection of 
$\gamma$-fold in the interval 11--20, as measured in the Innerball. 
The second spectrum, showed with filled squares, was obtained 
with the additional simultaneos selection  of clean and well resolved low 
energy $\gamma$-ray transitions of $^{42}$Ca in the Ge-detectors of 
the EUROBALL array. This particular 
residual nucleus was expected to be populated only in the decay from 
the highest spin region of the compound nucleus and, therefore, by gating 
on it we obtained data corresponding to the highest spins and free from fission 
contaminations.
The two spectra, normalized at $E_{\gamma}=$~6~MeV, show differences 
in the two regions: around $E_{\gamma}=$~9~MeV and around 
$E_{\gamma}=$~17~MeV, where evident yield excesses were found in the germanium 
gated spectrum. 
These excesses are better evidenced in the inset to Figure~\ref{fig:1}, 
which shows the ratio of the two measured spectra. 

\section{THE STATISTICAL MODEL ANALYSIS}

The statistical model analysis was made for the high energy $\gamma$-spectrum 
gated by the discrete transitions in $^{42}$Ca. 
This spectrum, being associated with a selected 
evaporation channel, was analyzed using the CASCADE code based 
on the Monte-Carlo technique.
This, in fact, allows to select only those decay chains which lead to 
the population of a specific residual nucleus.

The used Monte Carlo version of the CASCADE code, previously employed 
for the analysis of the heavier mass  data~\cite{FC_Hg,MK_Eu},
was implemented for this mass region by choosing the most appropriate values
of the statistical model parameters (e.g. level density was assumed to be 
in accordance with Reisdorf prescription~\cite{MKH87}, the yrast line was taken 
from the experiment and extrapolated by the liquid drop values).

\begin{figure}[htb]
\begin{center}
\includegraphics[scale=0.8]{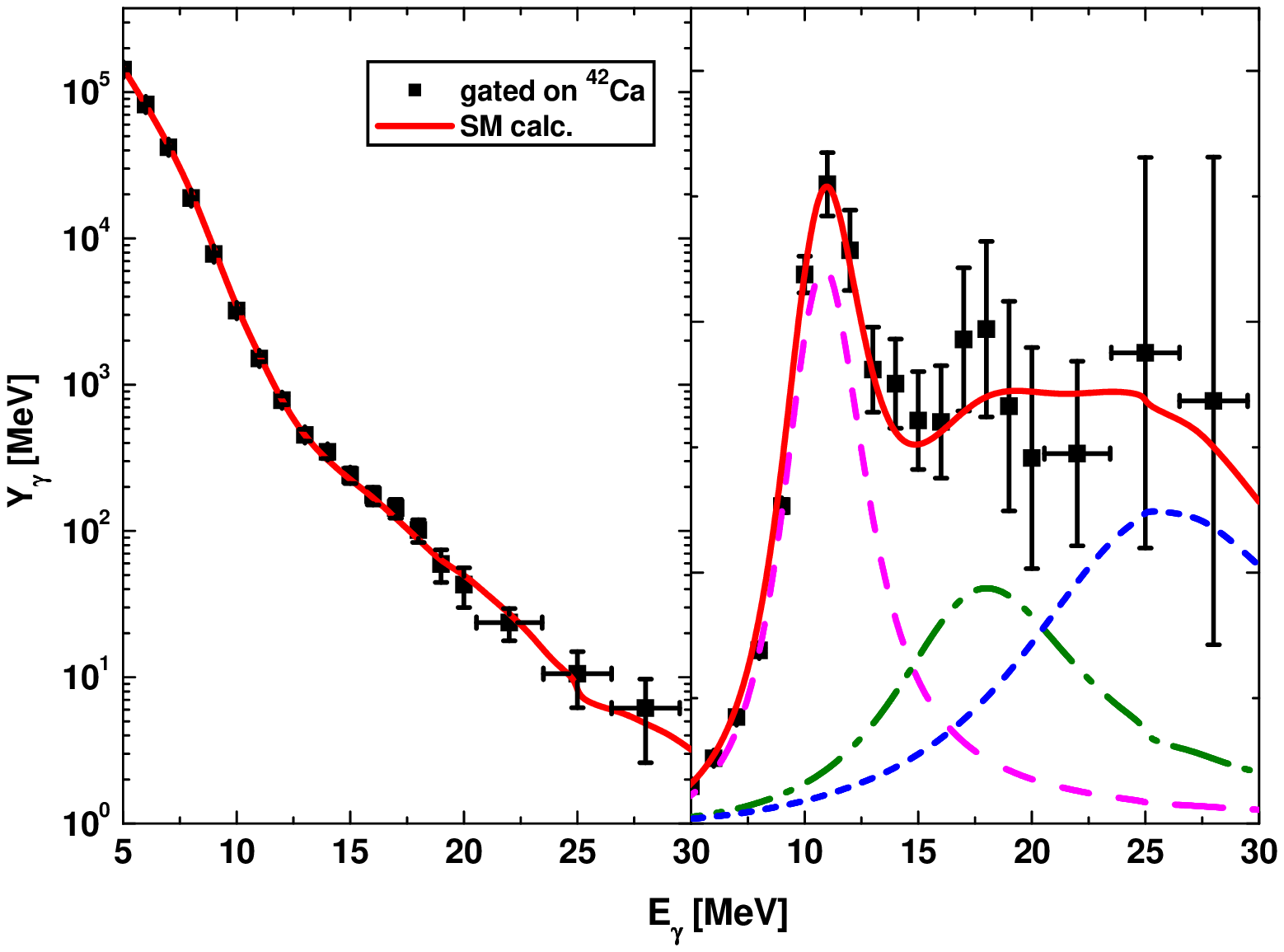}
\end{center}
\caption{
Left panel: The high-energy $\gamma$-ray spectrum gated by the $^{42}$Ca 
transitions and by high fold region (already shown in Fig.~\ref{fig:1}), 
in comparison with the best fitting statistical model calculations (full 
drawn line) assuming a 3-Lorentzian GDR lineshape with energies: 
$E_{GDR}=$~10.8(1), 18(1) and 26(2)~MeV; widths: $\Gamma_{GDR}=$~4.0(5), 10(1) 
and 16(3)~MeV; and strengths: $\sigma_{GDR}=$~0.35(5), 0.38(10) and 0.7(2).
Right panel: The deduced experimental GDR strength function (full drawn
line, see text for explanation) together with best fitting 3-Lorentzian 
function and its individual components.}
\label{fig:2}
\end{figure}

In the left panel of the Figure~\ref{fig:2}, the best fit to the experimental 
data is shown with the full line together with the data (filled squares).
It corresponds to a GDR  strength function consisting of 3 components 
at $E_{GDR}=$~10.8, 18 and 26~MeV 
exhausting approximately the entire EWSR strength. 
In particular, the need of a low energy component at~$\approx$~10~MeV
is consistent with the fact that this spectrum has an yield excess as 
compared to a more inclusive spectrum (see the inset to Figure~\ref{fig:1}).
The quality of the fit can be judged more clearly by inspecting 
the right panel of Fig.~\ref{fig:2}. 
This figure shows the GDR strength function data, represented by the quantity
$F_{3L}(E_{\gamma})*Y^{exp}_{\gamma}(E_{\gamma})/Y^{cal}_{\gamma}(E_{\gamma})$.
In that expression
$Y^{exp}_{\gamma}(E_{\gamma})$ and $Y^{cal}_{\gamma}(E_{\gamma})$
are the experimental and calculated spectra, respectively, shown 
in the left panel of Fig.~\ref{fig:2}.
The best fit GDR strength function, $F_{3L}(E_{\gamma})$,  consists 
of 3-component Lorentzian function 
with parameters (energy, width and strength) taken from the best fit 
to the experimental spectrum. $F_{3L}(E_{\gamma})$ and its individual 
components are also plotted in the figure. 
It should be noted, that the best fit line $F_{3L}(E_{\gamma})$ 
corresponds to the effective nuclear shape probed by the GDR oscillations.

\section{DISCUSSION}

The most striking feature of our results is the presence of the 
excess of the $\gamma$-yield 
at $\approx$~10~MeV, identified for the first time in this mass region. 
A possible explanation can be related to a mechanism such as the Coriolis 
splitting of the GDR lineshape controlled by the rotational frequency 
of the rotating nucleus. In fact, only in those
light nuclei one can reach the highest values of the rotational frequency 
(2--3~MeV in the spin interval 20--30~$\hbar$) and in this particular 
experiment we are able to select them much better than in the previous 
rather inclusive works. Therefore, we have investigated  in detail the effect 
of the Coriolis splitting in the GDR lineshape as a function of the spin 
(rotational frequency) and the quadrupole deformation type ($\gamma$) 
and the size ($\beta$). 

The expressions connecting the GDR vibrational frequencies to the rotational 
frequency~\cite{Nee,Gall} of the nucleus are reported in Table~\ref{table:1}.
\begin{table}[htb]
\caption{In this table we denote with $\omega$ the rotational frequency
         while $\omega_1$, $\omega_2$ and $\omega_3$ are the vibrational
	 frequencies along the axis in the intrinsic frame of the nucleus.
	 The expressions for the relative strengths of the 5 components
	 are taken from~\cite{JJG93}.}
\label{table:1}
\newcommand{\m}{\hphantom{$-$}}
\newcommand{\cc}[1]{\multicolumn{1}{c}{#1}}
\renewcommand{\tabcolsep}{3pc} 
\renewcommand{\arraystretch}{2.0} 
\begin{center}
\begin{tabular}{|c|c|c|}
\hline
\multicolumn{2}{|c|}{GDR rotational frequency}    & Strength \\
$\omega=0$    &   $\omega>0$            &          \\
\hline
$\omega_1$    &   $\omega_1$            & 
$\frac {1}{\omega_1}$\\
\hline
$\omega_2$    &   $\Omega_2 + \omega$   & 
$\frac {1}{\Omega_2} \frac {(\Omega_2-\omega)^2-0.5(\omega_2^2+\omega_3^2)}
{2\sqrt{\Delta}}$ \\
\cline {2-3}
              &   $\Omega_2 - \omega$   & 
$\frac {1}{\Omega_2} \frac {(\Omega_2+\omega)^2-0.5(\omega_2^2+\omega_3^2)}
{2\sqrt{\Delta}}$\\
\hline
$\omega_3$    &   $\Omega_3 + \omega$   & 
$\frac {1}{\Omega_3} \frac {-(\Omega_3-\omega)^2-0.5(\omega_2^2+\omega_3^2)}
{2\sqrt{\Delta}}$ \\
\cline {2-3}
              &   $\Omega_3 - \omega$   & 
$\frac {1}{\Omega_3} \frac {-(\Omega_3+\omega)^2-0.5(\omega_2^2+\omega_3^2)}
{2\sqrt{\Delta}}$ \\
\hline
\multicolumn{3}{|c|}
   {$\Omega_2=\frac {\omega_2^2+\omega_3^2}{2}+\omega^2+\sqrt{\Delta}$} \\
\multicolumn{3}{|c|}
   {$\Omega_3=\frac {\omega_2^2+\omega_3^2}{2}+\omega^2-\sqrt{\Delta}$} \\
\multicolumn{3}{|c|}
   {$\Delta=\frac{1}{4}(\omega_2^2-\omega_3^2)^2+2\omega^2(\omega_2^2+\omega_3^2)^2$} \\      
\hline
\end{tabular}\\[2pt]
\end{center}
\end{table}

\begin{figure}[htb]
\includegraphics[angle=-90,scale=0.6]{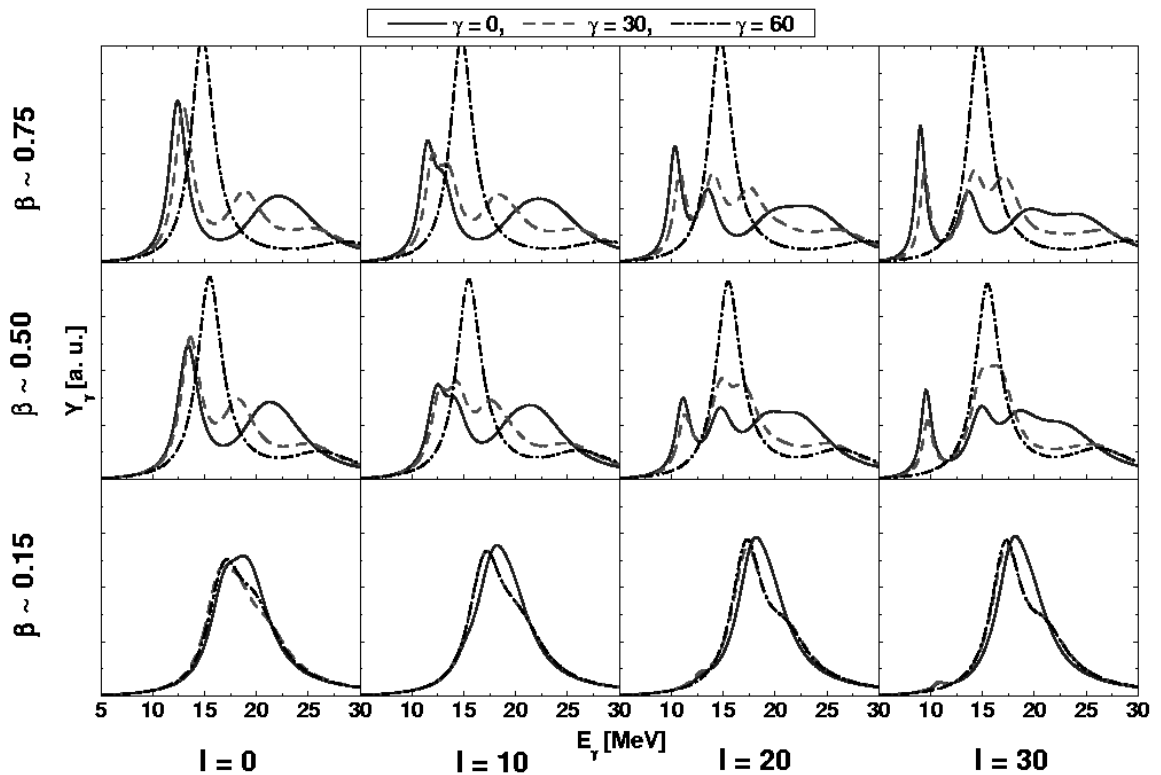}
\caption{
Calculated GDR lineshapes for $^{46}$Ti assuming different types 
of deformation (prolate $\gamma=$~0$^{\circ}$, full drawn line; triaxial 
$\gamma=$~30$^{\circ}$, long-dashed line; $\gamma=$~60$^{\circ}$ oblate, 
dashed-dotted line)
for different values of spin and quadrupole deformation size. This is to
illustrate the effect of the Coriolis splitting giving rise,
at highest rotational frequencies and prolate and triaxial deforamtions,
to a component around 10~MeV.}
\label{fig:3}
\end{figure}

Figure~\ref{fig:3} shows the calculated GDR lineshapes for $^{46}$Ti 
to emphasize how they evolve as the functions of $I$ and $\beta$ for the cases 
of prolate, oblate and triaxial shapes. One can clearly note that 
the Coriolis splitting is not affecting the oblate shapes 
($\gamma=$~60$^{\circ}$). In contrast, it produces strong effects, 
increasing as a function of spin, in the case of triaxial 
($\gamma=$~30$^{\circ}$) and especially prolate shapes ($\gamma=$~0$^{\circ}$). 
In particular, it produces a rather strong and narow peak at $\approx$~10~MeV 
for prolate and triaxial deformations with $\beta \geq$~0.4 which is similar 
to that we have observed in our experimental data.

In order to interprete the obtained GDR strength function data in the entire 
measured $\gamma$-ray energy region (5--30~MeV),
we use the same approach as has been adopted in almost all experiments 
concerning the GDR in hot and rotating nuclei~\cite{BBB,JJG,SNO}. 
This is based on the termal shape fluctuation model which predicts 
the GDR strength function corresponding to the shape ensemble probed by  
this collective excitation mode.
 
In the present case we have obtained the shape ensemble distributions 
(proportional to the the Boltzmann factor $exp(-F/T)$, where $F$ is 
the free energy and $T$ is the temperature)
 for the relevant total excitation energy (85~MeV)
and  for different spins using the new  parametrization of the liquid drop 
model LSD~\cite{Dudek}.

\begin{figure}[htb]
\begin{minipage}[t]{95mm}
\includegraphics[scale=0.68]{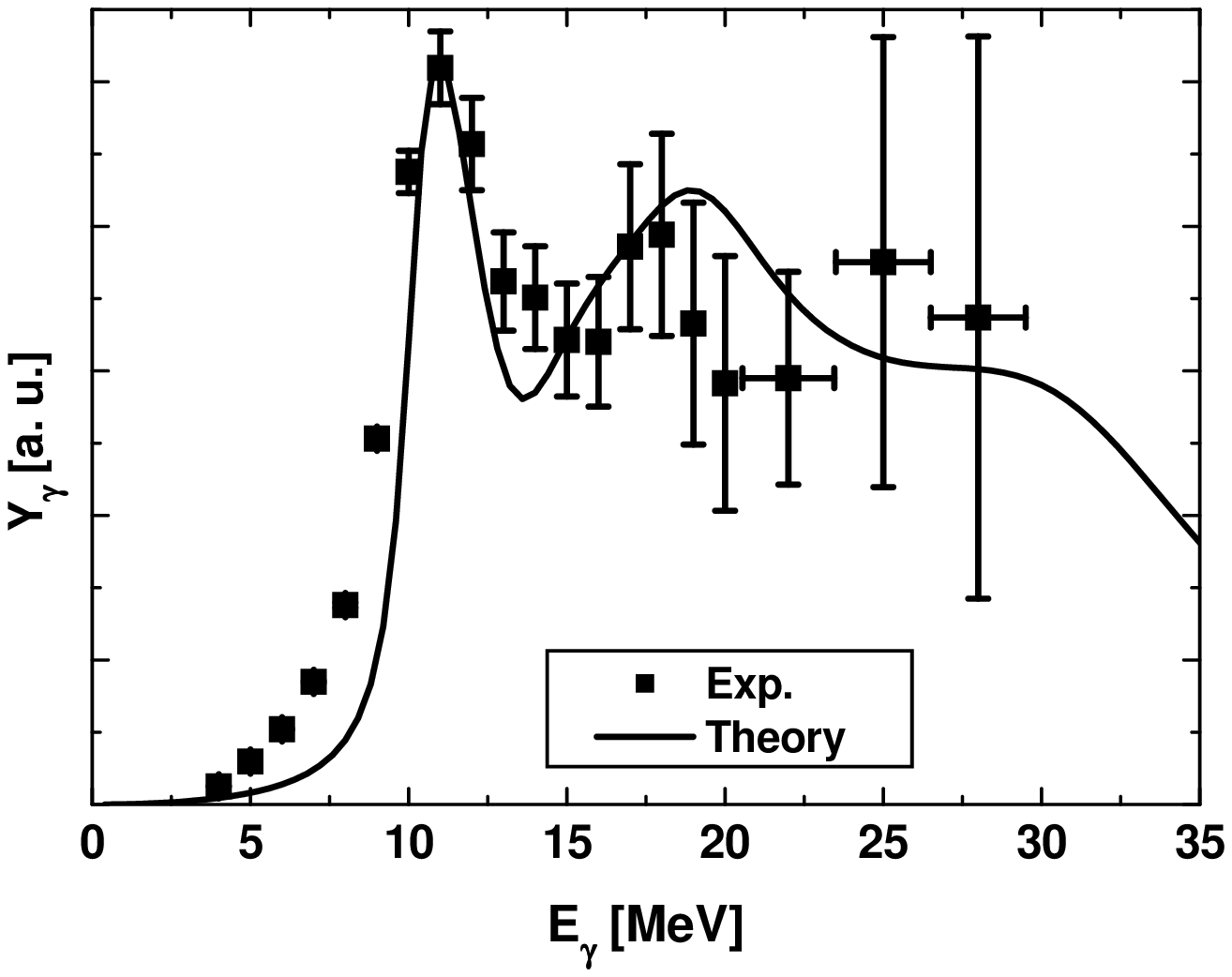}
\end{minipage}
\begin{minipage}[b]{62mm}
\caption{
The full drawn line shows the theoretical prediction 
(at $<T>$~$\approx$~2~MeV and in the spin region 28--34~$\hbar$) of the GDR 
lineshape in $^{46}$Ti obtained from the thermal shape fluctuation model 
based on free energies from the LSD model calculations~\cite{Dudek}. 
The filled squares are the experimental data shown also in the right panel 
of Figure~\ref{fig:2}.}
\label{fig:4}
\end{minipage}
\end{figure}

Those calculations show that the equilibrium shape of $^{46}$Ti undergoes 
the Jacobi transition from an oblate to a triaxial and a prolate shape 
for $I=$~28~$\hbar$. 
For the comparison with the experimental data at the highest 
spins populated in the reaction,
we select the calulations for the spin interval 28--34~$\hbar$ 
(this corresponds to an average temperature $T=$~2~MeV and rotational 
frequency $\omega=$~2.8~MeV). 
The calculations neglect shell effects, which are not expected 
to be important in this temperature region. The resulting GDR strength 
function, obtained as a weighted average (with the weight given by 
the Boltzmann factor) of strength functions calculated at each 
$\beta,\gamma$ point (including also Coriolis effects), is displayed 
with the full line in Fig.~\ref{fig:4} and compared with 
the experimental data.
The remarkable good agreement between the theoretical predictions 
and the present experimental results is very much in favour of 
the presence of the Jacobi transition. More evidence for this transition 
is also given 
by the fact that the peak at  $\approx$~10~MeV can originate only 
in the presence of a fast rotation in nuclei of prolate or triaxial shapes 
(see discussion at the beginning of the section).
In fact, for oblate shapes, typical for the equilibrium deformation 
at rotational frequencies lower than the critical value for the Jacobi 
transition, the Coriolis splitting is always absent.

A further signature for the Jacobi transition is the presence of two other 
broad components in the strength function at higher energies.
One component, at $\approx$~17~MeV, is clearly seen in the present data, 
while the other which is very broad (at 20--30~MeV) can only barely be
identified because of the very limited statistics at these high energies
in the present very exclusive spectrum.

In summary, the present work on the GDR $\gamma$-decay in the hot rotating 
nucleus, $^{46}$Ti, shows evidence for the expected Jacobi shape transition. 
It is based on the observation of two particular features in the measured 
GDR strength function. The first, already found in the previous 
works~\cite{MKH,Maj_Osaka}, 
is the presence of a high energy component related to large 
deformations. 
The second feature, identified for the first time in the present experiment, 
is the aperance of a GDR component at $\approx$~10~MeV (in region where 
the statistics is very high), which is interpreted as due to the Coriolis 
splitting of the lowest vibrational frequency (which corresponds 
to the dipole vibration along the long axis of the well deformed 
prolate or triaxial shape) and consequently shifting down a part 
of the strength. This shows the importance of the selection of high
rotational frequencies and of a good tagging on fusion-evaporation events
in the investigations of nuclear shapes through the GDR $\gamma$-decay.

This work was supported by the Polish State Committee for Scientific 
Research (KBN Grant No.~P03B~118~22), the European Commission contract EUROVIV,
Danish Research Council and the Italian INFN.

\end{document}